\documentclass[prd, preprint, nofootinbib]{revtex4}
\usepackage[dvipdfmx]{graphicx}
\usepackage{amsfonts}
\usepackage{latexsym}
\usepackage{amssymb}
\usepackage{epsfig}
\usepackage{color}
\usepackage{comment}
\usepackage{here}	
\usepackage{amsmath}

\newcommand{\im}[1]{\text{Im}\,#1}
\newcommand{\re}[1]{\text{Re}\,#1}

\begin{document}

\title{Modular symmetry in magnetized/intersecting D-brane models}

\author{$^1$Tatsuo Kobayashi, $^1$Satoshi Nagamoto, and  $^2$Shohei Uemura}
 \affiliation{
$^1$Department of Physics, Hokkaido University, Sapporo 060-0810, Japan\\
$^2$Department of Physics, Kyoto University, Kyoto 606-8502, Japan}



\begin{abstract}
We study the modular symmetry in four-dimensional low-energy effective field theory, 
which is derived from type IIB magnetized D-brane models and type IIA intersecting D-brane models. 
We analyze modular symmetric behaviors of perturbative terms and non-perturbative terms induced 
by D-brane instanton effects.
Anomalies are also investigated and such an analysis on anomalies suggests corrections in 
effective field theory.
\end{abstract}

\preprint{EPHOU-16-011,KUNS-2636}
\preprint{}

\vspace*{3cm}
\maketitle



\section{Introduction}

T-duality in string theory relates a theory with the compact space size $R$ to 
another theory with the size $1/R$.
Thus, T-duality is a quite non-trivial symmetry in string theory.
Indeed, one type of superstring theory is related  to 
different type of superstring theory by T-duality. 
(See for review \cite{Polchinski:1998rq}.)

T-duality has also a remnant in four-dimensional (4D) low energy effective field theory derived from 
superstring theory.
In particular, 4D low energy effective field theory of heterotic string theory 
with certain compactification is invariant under 
the modular transformation of the moduli $\tau$, 
\begin{equation}
\label{eq:modular}
\tau \rightarrow \frac{a\tau +b}{c\tau +d},
\end{equation}
with $ad-bc =1$ and $a,b,c,d \in {\bf Z}$, 
at least at the perturbative level.
This is the symmetry inside a 4D effective field theory, but not between two theories.
We refer this symmetry inside one effective field theory as the modular symmetry
in order to distinguish this symmetry from the T-duality between two theories.

The modular symmetry plays an important role in studies on 4D low energy effective field theory of 
heterotic string theory.
For example, moduli stabilization and supersymmetry breaking were studied with the assumption that 
non-perturbative effects are also modular invariant  \cite{Ferrara:1990ei,Cvetic:1991qm}.
Moreover, anomalies of this symmetry were analyzed \cite{Derendinger:1991hq,Ibanez:1992hc}.
The anomaly structure in heterotic string theory has a definite structure.\footnote{See also \cite{Araki:2007ss}.}
Their phenomenological applications were also studied (see e.g. \cite{Ibanez:1991zv,Kawabe:1994mj}).
In addition, modular invariant potential of the modulus was studied for cosmic inflation \cite{Kobayashi:2016mzg}.
Thus, the modular symmetry in 4D low energy effective field theory is important from 
several viewpoints such as theoretical one, particle physics and cosmology.

In this paper, we study the modular symmetry in 4D low-energy effective field theory derived from 
type II superstring theory.
In particular, we consider the 4D low-energy effective field theory derived from type IIB magnetized D-brane models 
and type IIA intersecting D-brane models.
Their 4D low-energy effective field theories have been studied (see for review \cite{Blumenhagen:2006ci,Ibanez:2012zz}).
We study the modular symmetry at perturbative level in their low-energy effective field theories.
The T-duality of Yukawa couplings between magnetized D-brane models and intersecting D-brane models 
was studied in \cite{Cremades:2004wa}.
That is very useful to our purpose.
We extend such analysis to show modular transformation of 4D low-energy effective field theory 
including 3-point and higher order couplings.
Also, their anomalies are examined and the anomaly structure could provide  non-trivial information 
like those in heterotic string theory.
Furthermore, we discuss non-perturbative effects.

The paper is organized as follows.
In section \ref{sec:perturbative}, we study the modular symmetry of Yukawa couplings and higher order couplings 
at the perturbative level 
in the 4D low-energy effective field theory derived from type IIB magnetized D-brane models. 
In section \ref{sec:supergravity-anomaly}, we study supergravity theory derived from 
type IIA intersecting D-brane models.
In particular, we investigate the anomaly structure of the modular symmetry.
In section \ref{sec:instanton}, we study the modular symmetry of non-perturbative terms induced by 
D-brane instanton effects.
Section \ref{sec:conclusion} is conclusion.

\section{Modular symmetry}
\label{sec:perturbative}

Here, we study the modular symmetry 
in the 4D low-energy effective field theory derived from type IIB magnetized D-brane models.

\subsection{Magnetized D-brane models}

We start with magnetized D9-brane models in type IIB theory.
We compactify six dimensional (6D) space to the 6D torus, e.g. three 2-tori.
The metric of the $r$-th 2-torus for $r=1,2,3$ is written by 
\begin{equation}
g =R^2_r \left(
\begin{array}{cc}
1 & {\rm Re}\tau_r  \\
{\rm Re}\tau_r & |\tau_r|^2
\end{array}
\right),
\end{equation}
on the real basis $(x_r,y_r)$, where 
$\tau_r$ denotes the complex structure modulus.
We denote the volume of the $r$-th 2-torus by 
${\cal A}_r=R^2_r{\rm Im}\tau_r$.
We use the complex coordinate $z_r=x_r+\tau_r y_r$.

\subsubsection{Yukawa couplings}

Here, we review analysis on Yukawa couplings in \cite{Cremades:2004wa}.
Our setup includes  several stacks of D9-branes with magnetic fluxes.
We assume that our setup preserves 4D N=1 supersymmetry.
Among several D-branes, we consider two stacks of $N_a$ and $N_b$ D9-branes,
which correspond to the $U(N_a)\times U(N_b)$ gauge symmetry.
We put  magnetic fluxes, $F^a_r(=F^a_{z_r\bar z_r})$ and  $F^b_r(=F^b_{z_r\bar z_r})$ on these D-branes along 
$U(1)_a$ and $U(1)_b$ directions of $U(N_a) = U(1)_a \times SU(N_a)$ and 
$U(N_b) = U(1)_b \times SU(N_b)$.
The magnetic fluxes must be quantized as $F^a_r=\frac{\pi i}{{\rm Im }\tau_r}m_a^r$ 
in the complex basis.
For simplicity, here we do not include Wilson lines \cite{Cremades:2004wa}.

The open strings between these magnetized branes have massless modes.
There appear $I_{ab}^r$ zero-modes on the $r$-th 2-torus, 
where  $I_{ab}^r=m_a^r - m_b^r$, and 
the total number of massless modes is given by their product, $I_{ab} = \prod_{r=1}^3 I_{ab}^r$.
Their zero-mode profiles on the $r$-th 2-torus are written by \cite{Cremades:2004wa}
\begin{equation}
\label{eq:basis1}
\psi^{j,N}(\tau_r, z_r) = \mathcal{N}_r \cdot e^{i \pi N z_r  {\rm Im}{z}_r/ \im{\tau}_r } \cdot \vartheta \left[
\begin{array}{c}
\frac{j}{N} \\
0
\end{array}
\right] ( N z_r, N \tau_r),\ 
\end{equation}
for $N=I_{ab}^r >0$, where $j$ denotes the zero-mode index 
for $j=1, \cdots, N$ (mod $N$), and $\mathcal{N}_r$ is the normalization factor given by 
\begin{equation}
\mathcal{N}_r = \left( \frac{2{\rm Im}{\tau}_r |N|}{\mathcal{A}^2_r} \right)^{1/4}.
\end{equation}
The $\vartheta$-function is defined as 
\begin{equation}
\vartheta \left[
\begin{array}{c}
a \\
b
\end{array}
\right] (\nu, \tau) = \sum_{l \in {\bf Z}} e^{\pi i (a+l)^2 \tau} e^{2 \pi i (a+l)(\nu+b)}.
\end{equation}

These zero-modes are also written by another basis,
\begin{equation}
\label{eq:basis2}
\chi^{j,N}(\tau_r,z_r) = \frac{\mathcal{N}_r}{\sqrt{N}} \cdot e^{i \pi N z_r  \im{z}_r/ \im{\tau}_r } \cdot \vartheta \left[
\begin{array}{c}
0 \\
\frac{j}{N}
\end{array}
\right] ( z_r, \tau_r/N ),\ \ \ \ \ j = 1,\cdots, N.
\end{equation}
These bases are related as
\begin{equation}
\chi^{j,N} = \frac{1}{\sqrt{N}} \sum_k e^{2\pi i \frac{jk}{N}} \psi^{k,N} .
\end{equation}

Note that the zero-mode profiles of bosonic and fermionic modes are the same in supersymmetric models.
For  $N=I_{ab}^r <0$, the zero-mode profiles are obtained by $\psi^{j,N}(\tau_r, z_r)^*$.

In addition to the above two stacks of D-branes, we consider another stack of $N_c$ D9-branes.
Then, there appear three types of massless modes, $a-b$, $b-c$, and $c-a$ modes.
Their Yukawa couplings among cannonically normalized fields can be obtained by 
overlap integral of wavefunctions, 
\begin{equation}
\label{eq:yukawa}
y_{ijk} = C_{abc} \, e^{\phi_{10}/2} \prod_{r=1}^3 \int dz_r d\bar{z}_r \ \psi^{i, {I}_{ab}^r}(z_r ) \cdot \psi^{j, {I}_{ca}^r}(z_r) \cdot \left( \psi^{k,{I}_{cb}^r}(z_r) \right)^*,
\end{equation}
where $C_{abc}$ is the moduli-independent coefficient and 
$\phi_{10}$ denotes the ten-dimensional dilaton.
Here, we set ${I}_{ab}^r + {I}_{ca}^r = -{I}_{bc}^r = {I}_{cb}^r$, because of gauge invariance.
To be exact, we should replace the zero-mode indexes $i,j,k$ by  $i^r,j^r,k^r$.
However, we denote them as  $i,j,k$ to simplify the equations.
Hereafter, we use a similar simplification.
In this computation, the following relation of zero-mode profiles, 
\begin{eqnarray}
\psi^{i,{I}_{ab}^r} \cdot \psi^{j,{I}^r_{ca}} &=& \mathcal{A}_r^{-1/2} (2 \im{\tau}_r)^{1/4} \left| \frac{{I}^r_{ab} {I}^r_{ca}}{{I}^r_{bc}} \right|^{1/4} \nonumber \\
&\cdot& \sum_m \psi^{i+j+{I}^r_{ab}m, {I}^r_{cb}}(z) \cdot \vartheta \left[
\begin{array}{c}
\frac{{I}^r_{ca}i - {I}^r_{ab} j + {I}^r_{ab} {I}^r_{ca} m}{- {I}^r_{ab} {I}^r_{bc} {I}^r_{ca}} \\
0
\end{array}
\right] \left( 0, \tau_r \left|{I}^r_{ab} {I}^r_{bc} {I}^r_{ca} \right|  \right) ,
\label{eq:phi-relation}
\end{eqnarray}
is very useful.
Then, the Yukawa coupling is written by \cite{Cremades:2004wa}
\begin{equation}
\label{eq:Yukawafull}
y_{ijk} = C_{abc}\, e^{\phi_{10}/2} \prod^3_{r=1} \left( \frac{2\im{\tau_{r}}}{{\mathcal{A}_{r}}^2} \right)^{1/4} \left| \frac{{I}^{r}_{1} {I}^{r}_{2}}{{I}^{r}_{1} + {I}^{r}_2} \right|^{1/4} \cdot  \vartheta \left[
\begin{array}{c}
\delta^{r}_{ijk} \\
0
\end{array}
\right] \left( 0, \tau_{r} \left|I^{r}_{ab} I^{r}_{bc} I^{r}_{ca} \right|  \right) ,
\end{equation}
where 
\begin{eqnarray*}
\delta_{ijk}^r &=& \frac{i}{{I}^r_{ab}} + \frac{j}{{I}^r_{ca}} + \frac{k}{{I}^r_{bc}}.
\end{eqnarray*}
Similarly, the Yukawa couplings can be written in the basis $\chi$,
\begin{eqnarray}
\label{eq:Yukawafull-chi}
y_{lmn} &=& C_{abc}\, e^{\phi_{10}/2} \prod^3_{r=1} \left( \frac{2\im{\tau_{r}}}{{\mathcal{A}_{r}}^2} \right)^{1/4} \left| \frac{{I}^{r}_{1} {I}^{r}_{2}}{{I}^{r}_{1} + {I}^{r}_2} \right|^{1/4}\cdot \left| I_{ab} I_{bc} I_{ca} \right|^{-1/2}  
\nonumber \\ 
& & 
\cdot  \vartheta \left[
\begin{array}{c}
0 \\
\delta^{r}_{ijk} 
\end{array}
\right] \left( 0, \tau_{r}/ \left|I^{r}_{ab} I^{r}_{bc} I^{r}_{ca} \right|  \right).
\end{eqnarray}

It would be convenient to use the 4D dilaton, 
\begin{equation}
e^{\phi_4} = e^{\phi_{10}} \prod_{r=1}^{3} ({\cal A}_{r})^{-1/2},
\end{equation}
and we define $\tilde I^{r}={ I^{r}}/{{\mathcal A}_{r}}$.
Then,  we can write the Yukawa coupling 
\begin{equation}
\label{eq:Yukawafull-4D}
y_{ijk} = C_{abc}\, e^{\phi_{4}/2} \prod^3_{r=1} \left({2\im{\tau_{r}}} \right)^{1/4} \left| \frac{\tilde{I}^{r}_{1} \tilde{I}^{r}_{2}}{\tilde{I}^{r}_{1} + \tilde{I}^{r}_2} \right|^{1/4} \cdot  \vartheta \left[
\begin{array}{c}
\delta^{r}_{ijk} \\
0
\end{array}
\right] \left( 0, \tau_{r} \left|I^{r}_{ab} I^{r}_{bc} I^{r}_{ca} \right|  \right).
\end{equation}

\subsubsection{Modular symmetry}

Now, let us study the modular transformation of the complex structure moduli $\tau_r$.
Recall that we use the basis, that the fields are normalized cannonically.
Thus, we just investigate the modular transformation of the Yukawa couplings.
The modular transformation (\ref{eq:modular})
is generated by the two generators, $s$ and $t$,
\begin{equation}
s: \tau \to -\frac{1}{\tau}, \qquad  t: \tau \to \tau + 1. 
\end{equation}
The modular function satisfies 
\begin{equation}
f(-1/\tau) = \tau^n f(\tau),
\end{equation}
where $n$ is called its modular weight.
It is obvious that ${\rm Im \tau}$ is invariant under $t$.
Under $s$, we have 
\begin{equation}
{\rm Im \tau} \to   \frac{1}{|\tau|^2} {\im \tau}. 
\end{equation}
The   $\vartheta$-function $\vartheta {\tiny \left[ \begin{array}{c} a \\ b  \end{array}\right]}(0,\tau)$ is the modular function with the modular weight 1/2.

The $\vartheta$-function part in the Yukawa coupling is transformed under $s:\tau \rightarrow -1/\tau$,
\begin{equation}
\vartheta \left[
\begin{array}{c}
\delta_{ijk} \\
0
\end{array}
\right] \left( 0, \tau \left| I_{ab} I_{bc} I_{ca} \right| \right)
\to \vartheta \left[
\begin{array}{c}
\delta_{ijk} \\
0
\end{array}
\right] \left( 0, -{\left| I_{ab} I_{bc} I_{ca} \right|}/{\tau} \right)  \ .
\end{equation}
Furthermore, using the Poisson resummation formula, 
we find 
\begin{equation}
 \vartheta \left[
\begin{array}{c}
\delta_{ijk} \\
0
\end{array}
\right] \left( 0, -{\left| I_{ab} I_{bc} I_{ca} \right|}/{\tau} \right) 
=
(-i\tau)^{1/2} \left| I_{ab} I_{bc} I_{ca} \right|^{-1/2} \vartheta \left[
\begin{array}{c}
0 \\
\delta_{ijk} 
\end{array}
\right] \left( 0, {\tau}/\left| I_{ab} I_{bc} I_{ca} \right| \right) .
\end{equation}
Thus, the $\tau$ dependent part in the Yukawa coupling transforms under $s$
\begin{equation}
(\im{\tau})^{1/4} \cdot \vartheta \left[
\begin{array}{c}
\delta_{lmn} \\
0
\end{array}
\right] \left( 0, \tau \left| I_{ab} I_{bc} I_{ca} \right| \right) \to
(\im{\tau})^{1/4} \cdot \left| I_{ab} I_{bc} I_{ca} \right|^{-1/2} \cdot \vartheta \left[
\begin{array}{c}
0 \\
\delta_{lmn} 
\end{array}
\right] \left( 0  , \tau /\left|I_{ab} I_{bc} I_{ca} \right|  \right) .
\end{equation}
This is nothing but the $\tau$ dependent part of the Yukawa coupling 
in the $\chi$ basis.
Therefore, the Yukawa coupling terms in 4D low energy effective field theory 
are invariant under modular transformation including basis change.

The above results can be extended to the magnetic flux,
\begin{equation}
F_{z \bar z} = \frac{\pi i}{\im \tau}\left(
\begin{array}{ccc}
\frac{m_a}{n_a} {\bf 1}_{n_a} & & \\
 & \frac{m_a}{n_a} {\bf 1}_{n_b}  & \\
 & & \frac{m_a}{n_a} {\bf 1}_{n_c}  \\
\end{array}
\right) ,
\end{equation}
by replacing $I_{ab}$ as $I_{ab }= n_b m_a - n_a m_b$.

\subsubsection{Higher order couplings}

We can study higher order couplings in a similar way \cite{Abe:2009dr}.
For example, the 4-point coupling can be obtained by computing 
integral of zero-mode profiles,
\begin{equation}
C_{abcd} e^{\phi_{10}} \prod_{r=1}^3 \int dz_r d\bar{z}_r \ \psi^{i, I^r_{ab}}(z_r) \cdot \psi^{j, I^r_{bc}}(z_r) \cdot \psi^{k, I^r_{cd}}(z_r) \cdot \left( \psi^{l, I^r_{ad}}(z_r) \right)^* .
\end{equation}
We use the relation (\ref{eq:phi-relation}), and then we obtain \cite{Abe:2009dr}
\begin{eqnarray}
y_{ijk\bar{l}} &=& C e^{\phi_{10}} \prod_{r=1}^3\left( \frac{2 \im{\tau}_r}{\mathcal{A}^2_r} \right)^{\frac{2}{4}} \left| \frac{I^r_{ab} I^r_{bc}}{M^r} \right|^{\frac{1}{4}} \cdot \left| \frac{M^r I^r_{cd}}{I^r_{ad}} \right|^{\frac{1}{4}} \\
&\ & \sum_{m\in {\bf Z}_{I^r_{ab}+I^r_{bc}}} \vartheta \left[
\begin{array}{c}
\frac{I^r_{bc} i -I^r_{ab} j + I^r_{ab} I^r_{bc} m}{I^r_{ab} I^r_{bc} M^r} \\
0
\end{array}
\right] \left( 0 , \tau_r I^r_{ab} I^r_{bc} M^r \right) \cdot  \vartheta \left[
\begin{array}{c}
\frac{I^r_{cd} l -I^r_{ad} k + I^r_{cd} I^r_{ad} r}{I^r_{cd} I^r_{ad} M^r} \\
0
\end{array}
\right] \left( 0 , \tau_r I^r_{cd} I^r_{ad} M^r \right), \nonumber
\end{eqnarray}
where $M^r=I^r_{ab} + I^r_{bc}$ and $i+j+k+I_{ab}^rm +(I^r_{ab}+I_{bc}^r)n = \ell + kI_{ad}r$ with a certain integer $n$.

Similarly, we can compute generic n-point couplings \cite{Abe:2009dr}, whose 
$\tau$ dependence as well as $\phi_4$ dependence appears in the form,
\begin{equation}
\label{eq:n-point}
e^{(n-2)\phi_4/2}\prod_{i=1}^{n-2} \prod_{r=1}^3({\rm Im}\tau_r )^{1/4} \cdot \vartheta \left[
\begin{array}{c}
{\delta_i^r} \\
0
\end{array}
\right] \left( 0 , \tau_r \alpha^r_i \right),
\end{equation}
for proper values of $\delta^r_i$ and $\alpha^r_i$, 
because we use the relation (\ref{eq:phi-relation}).
Note that the $\vartheta$-function multiplied $\im \tau^{-1/4}$ 
is invariant under modular transformation.
Thus, 4D low-energy effective field theory is 
invariant at perturbative level under modular transformation 
of the complex structure moduli, up to change of field basis.

Similarly, we can study the orientifold and orbifold compactifications.
For example, the zero-mode profiles on the $Z_2$ orbifold can be written by linear 
combinations of zero-mode profiles   on the torus \cite{Abe:2008fi}, 
\begin{equation}
\label{eq:wf-orbifold}
\psi^{j,N}(z)_{\rm orbifold} = \frac{1}{\sqrt 2}(\psi^{j,N}(z) + \psi^{N-j,N}(z)).
\end{equation}
Thus, the Yukawa couplings on the orbifold as well as higher order couplings can be written by 
linear combinations of Yukawa couplings on the torus \cite{Abe:2008fi}.
Then, the Yukawa couplings on the orbifold are also modular invariant in the same way as 
those on the torus.
Furthermore, the modular symmetry in magnetized D5 and D7-brane models 
can be studied in a similar way.
 
\section{Supergravity and anomaly}
\label{sec:supergravity-anomaly}

In this section, we study modular symmetry within the framework of string-derived supergravity 
and investigate its anomaly.

\subsection{Intersecting D-brane models}

In the previous section, we have studied modular symmetry in 
4D low energy effective field theory of magnetized D-brane models for 
cannonically normalized fields.
Here, we study type IIA intersecting D-brane models, which are T-dual to 
magnetized D-brane models.
In intersecting D-brane models, K\"ahler metric of matter fields was 
computed \cite{Blumenhagen:2006ci,Kors:2003wf,Lust:2004cx,Akerblom:2007uc,Blumenhagen:2007ip}.
In this section, we study the modular symmetry from the 
viewpoint of supergravity derived from intersecting D-brane models.
In particular, we study intersecting D6-brane models, where 
two sets of D6-branes, e.g. D6$_a$ and D6$_b$, intersect each other 
at the angle $\pi \theta_{ab}^r$ on the $r$-th 2-torus.

First, we write the supergravity fields in type IIB theory as
\begin{equation}
\re S = e^{-\phi_{10}} \prod_{r=1}^3 {\cal A}_{r}, \qquad \re T_{r} =  e^{-\phi_{10}} {\cal A}_{r}, \qquad 
U_{r} = i\tau_{r},
\end{equation}
where the imaginary parts of $S$ and $T_{r}$ correspond to certain axion fields.
Their K\"ahler potential is written by 
\begin{equation}
K = -\ln (S + \bar S) - \sum_{r=1}^3 \ln (T_{r} + \bar T_{r}) - \sum_{r=1}^3 \ln (U_{r} + \bar U_{r}) .
\end{equation}

We take T-dual along $x_r$ direction on each 2-torus from magnetized 
D9-branes to intersecting D6-branes.
Then, we replace 
\begin{equation}
T_{r} \longleftrightarrow U_{r}.
\end{equation}

We have seen that low-energy effective field theory of cannonically normalized fields
is modular symmetric for $\tau_r$ in type IIB magnetized D-brane models .
Thus, the low-energy effective field theory of type IIA intersecting D-brane models 
must have the symmetry under the modular transformation,
\begin{equation}
\label{eq:modular-T}
T_r \to \frac{a_rT_r - ib_r}{ic_rT_r + d_r},\ \qquad  a_r,b_r,c_r,d_r \in {\bf Z},\  \qquad a_rd_r-b_rc_r=1 ,
\end{equation}
both in canonically normalized field basis and in supergravity basis.

We take the T-dual of the Yukawa coupling (\ref{eq:Yukawafull-4D}) of magnetized D9-brane models, 
and then we can write the Yukawa coupling of intersecting D-brane models, 
\begin{equation}
\label{eq:Yukawafull-4D=inter}
y_{ijk} = C_{abc}\, e^{\phi_{4}/2} \prod^3_{r=1} \left({2\re{T_{r}}} \right)^{1/4} \left| \frac{\tilde{I}^{r}_{1} \tilde{I}^{r}_{2}}{\tilde{I}^{r}_{1} + \tilde{I}^{r}_2} \right|^{1/4} \cdot  \vartheta \left[
\begin{array}{c}
\delta^{r}_{ijk} \\
0
\end{array}
\right] \left( 0, T_{r} \left|I^{r}_{ab} I^{r}_{bc} I^{r}_{ca} \right|  \right) ,
\end{equation}
where 
\begin{equation}
e^{\phi_4} = \frac{(\re U_{1}U_{2}U_{3})^{1/2}}{\re S}.
\end{equation}

Within the framework of supergravity, physical Yukawa couplings are written by 
\begin{equation}
\label{eq:Yukawasugra}
y_{ijk} = \left( K_{ab} K_{bc} K_{ca} \right)^{-1/2} e^{K/2} W_{ijk} \ ,
\end{equation}
where $W_{ijk}$ denotes the holomorphic Yukawa coupling in the superpotential, i.e., 
\begin{equation}
{W} = W_{ijk} \Phi_i \Phi_j \Phi_k + \cdots \ ,
\end{equation}
$K$ is the K\"ahler potential, and $K_{ab}$, $K_{bc}$, $K_{ca}$ are 
the K\"ahler metric of the ab, bc, ca sectors, respectively.
Then, the relation (\ref{eq:Yukawasugra}) requires that 
\begin{equation}
\label{eq:KKK-relation}
K_{ab} K_{bc} K_{ca} \propto \prod_r (T_{r}+ \bar T_{r})^{-3/2} \ .
\end{equation}

The  K\"ahler metric of matter fields was 
computed \cite{Blumenhagen:2006ci,Kors:2003wf,Lust:2004cx,Akerblom:2007uc,Blumenhagen:2007ip}.
The K\"ahler metric of the ab sector would be written as 
\begin{equation}
\label{eq:Kahler-metric-1}
K_{ab} = \prod_r (T_{r}+ \bar T_{r})^{\nu(\theta^{r}_{ab})} \ .
\end{equation}
For example, in Refs. \cite{Lust:2004cx,Akerblom:2007uc,Blumenhagen:2007ip}, the following Ansatz,
 \begin{equation}
 \label{eq:Kahler-metric-2}
 \nu (\theta^{r}_{ab}) = - \frac12 \pm \frac12 {\rm sign}(I_{ab})\theta_{ab}^{r} \ ,
 \end{equation}
was discussed by comparing the holomorphic and physical gauge couplings and 
threshold corrections.
They satisfy the above relation (\ref{eq:KKK-relation}) when 
\begin{equation}
{\rm sign}(I_{ab})\theta_{ab}^{r} + {\rm sign}(I_{bc})\theta_{bc}^{r} +{\rm sign}(I_{ca})\theta_{ca}^{r} = 0 \ .
\end{equation}

Similarly, the n-point couplings in magnetized D-brane models include the $\tau$ dependent factor (\ref{eq:n-point}).
Then, its T-dual intersecting D-brane models include $(2\re T_{r})^{n-2}/4$.
That requires that the product of the K\"ahler metric satisfies
\begin{equation}
 K_{a_1a_2} K_{a_2 a_3} \cdots K_{a_n a_1} =  \prod_r (T _{r} + \bar T_{r})^{-n/2}.
\end{equation}
This relation is also satisfied by (\ref{eq:Kahler-metric-2}) when 

\begin{equation}
{\rm sign}(I_{a_1a_2})\theta_{a_1a_2}^{r} +{\rm sign}(I_{a_2a_3})\theta_{a_2a_3}^{r} + \cdots 
+ {\rm sign}(I_{a_na_1})\theta_{a_na_1}^{r}   = 0 .
\end{equation}

We can take the T-dual of type IIA intersecting D-brane models along the $y_r$ direction, 
\begin{equation}
{\rm type~IIB~model~X} \underset{{\rm T-dual~along~} x_r}{\Longleftrightarrow} {\rm type~IIA~model} \underset{{\rm T-dual~along~} y_r}{\Longleftrightarrow} {\rm type~IIB~model~Y}
\end{equation}
%
and then obtain type IIB magnetized D-brane models, which are different from one discussed in the previous section, 
The relation between these two type IIB models was studied in \cite{Cremades:2004wa}, in particular Yukawa couplings.
Our results in the precious section can be understood as such two different theories through double T-duality 
such as  \cite{Cremades:2004wa}, 
but in any rate we are interested in the modular symmetry in one 4D low-energy effective field theory 
as mentioned in Introduction.

\subsection{Anomaly}

In the previous section, the modular symmetry in the supergravity basis was studied.
The chiral multiplet, $\Phi_{ab}$ in the ab sector  has the 
K\"ahler metric (\ref{eq:Kahler-metric-1}).
Thus, the chiral multiplet, $\Phi_{ab}$,  transforms 
\begin{equation}
\Phi_{ab} \rightarrow (ic_rT_r+d_r)^{-\nu(\theta_{ab}^r)} \Phi_{ab} \ ,
\end{equation}
under the modular transformation (\ref{eq:modular-T}).
That is, the matter field has the modular weight $\nu(\theta_{ab}^{r})$ under 
the modular transformation of the $r$-th 2-torus.

Such a modular transformation may be anomalous.
The supergravity Lagrangian includes the following couplings,
\begin{equation}
\left (\frac12 \re f \bar \lambda \gamma^\mu \lambda - \frac12 K_{i\bar{j}}\bar \psi_{j} \gamma^\mu \psi_i \right) \frac12
V^{\rm K\ddot ahler}_\mu + \left( \frac12 K_{i\bar{j}}\bar \psi_{j} \gamma^{\mu} \psi_l (-i \Gamma_{ikl} \partial_\mu \psi_k) +h.c. \right),
\end{equation}
where $\lambda$ denotes the gaugino,  $K_{ii}$ is the K\"ahler metric of $\Phi_i$ with the bosonic and fermionic components, $\phi_i$ and $\psi_i$,
\begin{equation}
\Gamma_{ij k} = \frac{\partial}{\partial \phi^i} \ln K_{jk}, \qquad 
V^{\rm K\ddot ahler}_\mu = -i \left( \frac{\partial K}{\partial \phi_i}\partial_\mu \phi_{i} - 
\frac{\partial K}{\partial \bar \phi_{j}}\partial_\mu \bar \phi_j \right) \ .
\end{equation}

These couplings induce the anomaly of modular symmetry.
Its anomaly coefficient of mixed anomaly with the $SU(N_a)$ gauge group is written by  \cite{Derendinger:1991hq}
\begin{equation}
A^r_a = - C_2(G_a) + \sum_{{\rm matter},b}T(R_a)(1+2\nu(\theta^{r}_{ab})),
\end{equation}
where $C_2(G_a)$ is the quadratic Casimir and $T(R_a)$ is the Dynkin index of 
the representation $R_a$.
For simplicity, we consider the intersecting D-brane models on torus.
In this case, we can write
\begin{equation}
A^r_a = - N_a + \frac12 \sum_{b}N_bI_{ab}(1+2\nu(\theta^{r}_{ab})).
\end{equation}

This anomaly can be cancelled by two ways \cite{Derendinger:1991hq,Ibanez:1992hc}.
One is moduli dependent threshold corrections and 
another is generalized Green-Schwarz mechanism.
The latter would lead to mixing of moduli, e.g. in K\"ahler potential.
In order to see it, we first review briefly on anomalous U(1) and Green-Schwarz mechanism 
in the next subsection \cite{Aldazabal:2000dg,Blumenhagen:2006ci,Ibanez:2012zz}.

\subsubsection{Anomalous U(1)}

First, let us consider the D6$_b$-branes wrapping the 3-cycle 
$[\Pi_b]$, whose wrapping numbers are $(n^r_b,m^r_b)$ along $(x_r,y_r)$.
We introduce the basis of 3-cycles, $[\alpha^0]$ and $[\alpha^k]$ with $k=1,2,3$, 
such that $[\alpha^0]$ is  along $(1,0)$ for all of $(x_r,y_r)$, 
while $[\alpha^k]$ is along $(1,0)$ only for $r=k$ and $(0,1)$ for the others.
We also introduce their duals $[\beta^k]$ such that $[\alpha^i]\cdot[\beta^k]=\delta_{ik}$.
These D6-branes correspond to $U(N_b)$ gauge group, and its gauge kinetic function $f_b$ is written by 
\begin{equation}\label{eq:f-tree}
f_b = q_b^0S - q^r_b U_r,
\end{equation}
where 
\begin{equation}
q_b^0= [\Pi_b]\cdot[\beta^0]=n_b^1n_b^2n_b^3, \qquad 
q_b^i= [\Pi_b]\cdot[\beta^i]=n_b^im_b^jm_b^k,
\end{equation}
where $i\neq j \neq k \neq i$.

Now, we study the $U(1)_a-SU(N_b)^2$ mixed anomaly.
Its anomaly coefficient can be written by 
\begin{equation}
N_a I_{ab} = q^0_bQ^0_a + \sum_i   q^i_bQ^i_a,  
\end{equation}
where 
\begin{equation}
Q_b^0= [\Pi_b]\cdot[\alpha^0], \qquad 
Q_b^i= [\Pi_b]\cdot[\alpha^i].
\end{equation}
This anomaly can be cancelled by the shift of moduli,
\begin{equation}
S \rightarrow S + Q_a^0 \Lambda_a, \qquad U_r - Q_a^r\Lambda_a,
\end{equation}
in the gauge kinetic function $f_b$ under the $U(1)$ transformation,
\begin{equation}
V_a \rightarrow V_a + \Lambda_a + \bar \Lambda_a.
\end{equation} 
This means that the K\"ahler potential is not invariant, but the following 
K\"ahler potential is invariant,
\begin{equation}
K = -\ln (S+\bar S - Q_a^0 V_v) - \sum_r \ln (U_r + \bar U_r - Q_a^r V_a)- \sum_ i \ln (T_r + \bar T_r).
\end{equation}

The Green-Schwarz mechanism is the same in the toroidal, orientifold and orbifold compactifications.

\subsubsection{Anomaly cancellation of modular symmetry}

As mentioned above, the modular anomaly can be canceled by two ways \cite{Derendinger:1991hq,Ibanez:1992hc}.
One is moduli dependent threshold corrections and 
another is generalized Green-Schwarz mechanism.
In general, the gauge kinetic function has one-loop threshold corrections due to massive modes as
\begin{equation}
f_a^{\rm(one-loop)} = f_a + \sum_i \Delta_a(T_r),
\end{equation}
where the first term in the RHS corresponds to Eq.(\ref{eq:f-tree}).
The threshold corrections are computed explicitly \cite{Dixon:1990pc,Lust:2003ky,Akerblom:2007uc,Blumenhagen:2007ip} and its typical form is 
\begin{equation}
\Delta_a(T_r) = \frac{\tilde b}{4\pi^2} \ln [\eta(iT_r)],
\end{equation}
where $\tilde b$ is beta-function coefficient due to massive modes, 
and $\eta(iT)$ is the Dedekind eta function.
The Dedekind eta function has the modular weight $1/2$.
This threshold correction can cancel the anomaly partially.
The other part of anomaly can be canceled by the generalized Green-Schwarz mechanism, where 
we impose the following transformation
\begin{equation}\label{eq:GS-SU}
S \rightarrow \frac{1}{8\pi^2}\sum_{r}\delta_{GS}^r \ln(ic_rT_r+d_{r}), \qquad 
U_i \rightarrow \frac{-1}{8\pi^2}\sum_r\delta_{GS}^{r,i} \ln(ic_rT_r+d_r),
\end{equation}
under the modular transformation (\ref{eq:modular-T}).
That is, the generalized Green-Schwarz mechanism could cancel the anomaly proportional to 
\begin{equation}
q^0_a\delta_{GS}^r + \sum_i q^i_a \delta_{GS}^{r,i}.
\end{equation}
By comparison with the total anomaly as well as the U(1) anomaly, a plausible Ansatz would be,
\begin{equation}
\delta_{GS}^i =\sum_b Q_b^0(\nu(\theta_{ab}^{(i)}) + c), \qquad \delta_{GS}^{i,r} =\sum_b Q_b^r(\nu(\theta_{ab}^{(i)}) + c),
\end{equation} 
where $c$ is constant.
In this case, the coefficient $\tilde b$ may be obtained  
\begin{equation}
\tilde b = N_a -\frac12 \sum_bN_bI_{ab}(1-2c),
\end{equation}
to cancel the modular anomaly.
Indeed, the threshold correction, 
\begin{equation}
\Delta_a = \frac{N_a}{4\pi^2}\ln [ \eta(iT_i) ], 
\end{equation}
was discussed in \cite{Akerblom:2007uc,Blumenhagen:2007ip}.

The transformation (\ref{eq:GS-SU}) implies that 
K\"ahler potential is not invariant under the modular transformation.
The K\"ahler potential must be modified as 
\begin{equation}
K=-\ln (S+\bar S - \sum_i\frac{\delta_{GS}^i}{8 \pi^2}(T_i + \bar T_i)) -
\sum_j \ln (U_j + \bar U_j - \sum_i\frac{\delta_{GS}^{i,j}}{8 \pi^2}(T_i + \bar T_i)) - 
\sum_i \ln (T_i + \bar T_i). 
\end{equation}
That is, the moduli mix, and instead of $S$ and $U^i$, the linear combinations, 
\begin{equation}
S+\bar S - \sum_i\frac{\delta_{GS}^i}{8 \pi^2}(T_i + \bar T_i), \qquad 
U_j + \bar U_j - \sum_i\frac{\delta_{GS}^{i,j}}{8 \pi^2}(T_i + \bar T_i),
\end{equation}
must appear in 4D low-energy effective field theory.
Similar linear combinations were discussed in \cite{Blumenhagen:2007ip}, 
although linear combinations in \cite{Blumenhagen:2007ip} include mixture of all the moduli.\footnote{ 
The sigma model anomaly concerned about $U_i$ is also discussed in \cite{Blumenhagen:2007ip}.}

Here, we return back to the type IIB model studied in section \ref{sec:perturbative}
.
Similar to the above, we may need to replace,
\begin{equation}
\label{eq:moduli-mix}
S+ \bar S \rightarrow S+\bar S - \sum_i\frac{\delta_{GS}^i}{8 \pi^2}(U_i + \bar U_i), \qquad 
T_j + \bar T_j \rightarrow T_j + \bar T_j - \sum_i\frac{\delta_{GS}^{i,j}}{8 \pi^2}(U_i + \bar U_i),
\end{equation}
in 4D low-energy effective field theory.
For example, the 4D dilaton factor in the Yukawa coupling would be modified as
\begin{equation}
e^{\phi_4} \rightarrow  \frac12 \frac{\left( \prod_i T_j + \bar T_j - \sum_i\frac{\delta_{GS}^{i,j}}{8 \pi^2}(U_i + \bar U_i)   \right)^{1/2}}{S+\bar S - \sum_i\frac{\delta_{GS}^i}{8 \pi^2}(U_i + \bar U_i)}.
\end{equation}

\section{D-brane instanton effects}
\label{sec:instanton}

In section \ref{sec:perturbative}, we studied the modular symmetry of perturbative terms in 
Lagrangian.
In this section, we study terms due to non-perturbative effects, in particular 
terms induced by D-brane instanton effects.
First, we study an illustrating example, and then we will discuss generic aspects.

\subsection{Example}

In this subsection, we study a Majorana mass term induced by a E5-brane in Type IIB magnetized orientifold models with O9-planes compactified on $Z_2\times Z_2'$ torus .
In these models, the non-perturbative corrections to superpotential are written 
as \cite{Blumenhagen:2006xt,Blumenhagen:2009qh}\footnote{See for explicit computations on intersecting D-brane 
orbifold models, 
e.g. \cite{Kobayashi:2015siy}.}
\begin{equation}
\Delta W= \int d\alpha^1 \cdots d\alpha^n  e^{-S_{\rm int}} e^{-S} .
\label{eq:nonpert}
\end{equation}
In (\ref{eq:nonpert}), $\alpha^i$ denotes a fermionic zero-mode of the E5-brane and $S$ denotes the classical action of E5-brane.
$S_{\rm int}$ denotes interaction terms including fermionic zero-modes as 
\begin{equation}
S_{{\rm int}} \sim y_{i_1\cdots i_n,j_1\cdots j_m}\alpha^{i_1}\cdots \alpha^{i_n} \Phi_{j_1}\cdots \Phi_{j_m},
\end{equation}
where $y_{i_1\cdots i_n,j_1\cdots j_m}$ is a $(n+m)$-point coupling and $\Phi_{j}$ is the chiral superfield of the models.
Then, we can obtain a Majorana mass term if there are two fermionic zero-modes and 3-point couplings like $y_{ijk}\alpha^i \beta^j \Phi_k$.
The Majorana mass is generated as
\begin{equation}
M_s^2\int d^2\alpha d^2\beta e^{y_{ijk}\alpha^i \beta^j \Phi_k}=M_s^2\epsilon_{ij}\epsilon_{kl} y_{ikm}y_{jln} \Phi_m\Phi_n.
\label{eq:majorana}
\end{equation}
In this subsection, we concentrate on the $r$th 2-dimensional torus with two D-branes wrapping whole compact space for simplicity.
We put the magnetic fluxes $\frac{\im \tau}{\pi i}F_r^a=2$ on one D-brane and $\frac{\im \tau}{\pi i}F_r^b=-2$ on the other D-brane.
For simplicity, all Wilson lines are set to zero in this subsection too.
Then, there are three chiral fermions between these two branes.
These modes are given by the linear combinations of the wave functions on the covering torus $\psi^i$,
\begin{equation}
\psi^i(z,\bar{z})=\left( \frac{4\cdot 2{\rm Im}\tau }{\mathcal{A}^2} \right)^{1/4} e^{i\pi 4 z {\rm Im} z/{\rm Im}\tau}\vartheta \left[
\begin{array}{c}
i/4\\
0\\
\end{array}
\right](4z,4\tau),
\end{equation}
where $i\in\{0,1,2,3 \}$.
The three zero-modes on the orbifold are given by Eq.(\ref{eq:wf-orbifold}) \cite{Abe:2008fi}.
That is, two of them, $\Phi_0$ and $\Phi_2$ correspond to $\psi^0$ and $\psi^2$, respectively, while 
$\Phi_1$ is given by 
\begin{equation}
\frac{1}{\sqrt{2}} (\psi^1 + \psi^3). 
\end{equation}
In addition, a E5-brane with no magnetic flux induces  two zero-modes between the E-brane and the D-branes.
These zero-modes are given by
\begin{equation}
\alpha^j(z,\bar{z})=\left( \frac{2\cdot 2{\rm Im}\tau }{\mathcal{A}^2} \right)^{1/4} e^{i\pi 2 z {\rm Im} z/{\rm Im}\tau}\vartheta \left[
\begin{array}{c}
j/2\\
0\\
\end{array}
\right](2z,2\tau),
\end{equation}
\begin{equation}
\beta^k(z,\bar{z})=\left( -\frac{2\cdot 2|{\rm Im}\bar{\tau}| }{\mathcal{A}^2} \right)^{1/4} e^{i\pi 2 \bar{z} {\rm Im} \bar{z}/{\rm Im}\bar{\tau}}\vartheta \left[
\begin{array}{c}
k/2\\
0\\
\end{array}
\right](2\bar{z},-2\bar{\tau}).
\end{equation}
Then, Yukawa couplings are written by
\begin{equation}
\begin{split}
y_{ijk}
=  &\left( \frac{4|{\rm Im}\bar{\tau}| }{\mathcal{A}^2} \right)^{\frac{1}{2}}  \sum_{m=0}^3 
\vartheta \left[
\begin{array}{c}
\frac{2j-2k+4m}{16}\\
0\\
\end{array}
\right](0,-16\bar{\tau})
\int_{T^2} dzd\bar{z} \\
&\begin{cases}  
\left( \frac{4\cdot 2{\rm Im}\tau }{\mathcal{A}^2} \right)^{\frac{1}{4}} \vartheta \left[
\begin{array}{c}
\frac{i}{4}\\
0\\
\end{array}
\right](4z,4\tau) \vartheta \left[
\begin{array}{c}
\frac{j+k+2m}{4}\\
0\\
\end{array}
\right](-4\bar{z},-4\bar{\tau})  ~~~~~~~~~~~~~~~~~~~~~~~~~~~~~~~~~  i=0,2, \\ 
\frac{1}{\sqrt{2}}\left( \frac{4\cdot 2{\rm Im}\tau }{\mathcal{A}^2} \right)^{\frac{1}{4}} 
\left( 
\vartheta \left[
\begin{array}{c}
\frac{1}{4}\\
0\\
\end{array}
\right](4z,4\tau)+
\vartheta \left[
\begin{array}{c}
\frac{3}{4}\\
0\\
\end{array}
\right](4z,4\tau)
\right)\vartheta \left[
\begin{array}{c}
\frac{j+k+2m}{4}\\
0\\
\end{array}
\right](-4\bar{z},-4\bar{\tau}) \ \ i=1.
\end{cases}
\end{split}
\end{equation}
Complete 3-point couplings are products of 3-point couplings of those on each 2 dimensional torus and 10 dimensinal string coupling.
Majorana mass term is written as (\ref{eq:majorana}).
This Majorana mass term is invariant under the modular transformation of 
the complex structure moduli since its dependence on complex structure moduli is determined by that of perturbative 3-point couplings and it is invariant under the modular transformation.
The modular symmetry is not violated by the non-perturbative effects in this case.

\subsection{Generic discussion}

The example in the previous subsection shows the modular symmetry of non-perturbative terms induced by D-brane instanton effects 
for the complex structure moduli in type IIB magnetized D-brane models.
Moreover, this example suggests a generic aspect.
The D-brane instantons induce the non-perturbative terms such as 
\begin{equation}
Ce^{-Vol(E5)}\left( \prod_i y^{(n_i)}(\tau) \right) \Phi_1 \cdots \Phi_m,
\end{equation}
where $C$ is a moduli-independent coefficient\footnote{More precisely, the coefficient C may include a functional determinant of Dirac operator as well as bosonic Laplacian operator produced by the integration of massive modes 
\cite{Witten:1999eg,Blumenhagen:2006xt}. However, these coefficients are canceled if the SUSY is not broken.
Even if the SUSY is broken, eigenvalues of Dirac operator and Laplacian operator depend only on ${\cal A}^r$, but they are  independent of the complex structures  \cite{Cremades:2004wa}.
Thus our conclusion would not be affected by this coefficient.}.
Here, $Vol(E5)$ denotes the volume of D-brane instanton in the compact space, 
and it depends only on ${\cal A}^{r}$, but not $\tau$.
Furthermore, $y^{(n)}$ denote the couplings among zero-modes and 4D fields $\Phi_i$, 
and these are computed in the same way as perturbative couplings shown in section \ref{sec:perturbative}. 
The $\tau$ dependence appears only through these couplings $y^{(n)}$.
Therefore, terms induced by D-brane instanton effects are also modular symmetric.

In this section, we have not taken into account the moduli mixing so far.
However, the discussion in section III would lead to modification such as (\ref{eq:moduli-mix}).


\section{Conclusion}
\label{sec:conclusion}

We have studied the 4D low-energy effective field theory, 
which is derived from type IIB magnetized D-brane models 
and type IIA intersecting D-brane models.
We have studied modular symmetric behavior of perturbative terms.
Also, such analysis has been extended to 
non-perturbative terms induced by D-brane instanton effects.
We have also investigated the anomaly of the modular symmetry.
Its cancellation would require moduli mixing correction terms 
in low-energy effective field theory.
Thus, the modular symmetry is important to understand 
the 4D low-energy effective field theory of superstring theory.


\section*{Acknowledgments}
 T.K. and S.U. are supported in part by
the Grant-in-Aid for Scientific Research No.~26247042 and No.~15J02107 from the Ministry of Education,
Culture, Sports, Science and Technology  in Japan.

%




\end{document}